\documentclass[twocolumn,secnumarabic,amssymb, nobibnotes, aps, prl,superscriptaddress]{revtex4-1}
\usepackage{graphicx}
\usepackage{colortbl}

\definecolor{dg}{rgb}{0.8,0.8,0.8}
\definecolor{lg}{rgb}{0.9,0.9,0.9}
\definecolor{r}{rgb}{1.0,0.0,0.0}
\definecolor{b}{rgb}{0.0,0.0,1.0}
\newcommand{\tc}{\textcolor}

\begin{document}

\title{Utilizing gene regulatory information to speed \\ up the calculation of elementary flux modes}
\author{Christian Jungreuthmayer}
\email{christian.jungreuthmayer@acib.at}
\author{David E. Ruckerbauer}
\author{J\"urgen Zanghellini}
\email{juergen.zanghellini@acib.at}
\affiliation{Austrian Centre of Industrial Biotechnology, Vienna, Austria, EU}
\affiliation{\mbox{Department of Biotechnology, University of Natural Resources and Life Sciences, Vienna, Austria, EU}}

%
%
%

\begin{abstract}

Despite the significant progress made in recent years,
the computation of the complete set of elementary flux modes of large or even genome-scale metabolic networks is still impossible.
We introduce a novel approach to speed up the calculation of elementary flux modes
by including transcriptional regulatory information into the analysis of metabolic network.
Taking into account gene regulation dramatically reduces the solution space and
allows the presented algorithm to constantly eliminate biologically infeasible modes at an early stage of the computation procedure.
Thereby, the computational costs, such as runtime, memory usage and disk space are considerably reduced.
Consequently, using the presented mode elimination algorithm pushes the size of metabolic networks that can be studied by elementary flux modes to new limits.

\end{abstract}

\maketitle

\section{Introduction}

Elementary flux modes (EFM) are indivisible sets of reactions that represent biologically meaningful pathways \cite{schuster2000,schuster1999} under steady state condition.
Removing only a single reaction of an EFM results in the extinction of the entire pathway.
Consequently, EFMs can be used to decompose metabolic networks mathematically and investigate them unbiasedly.
For that reason EFMs have gained increasing attention in the field of metabolic engineering in recent years.
However, the computation of EFMs is of combinatorial complexity \cite{klamt2002}.
Hence, the computational costs for calculating EFMs increase sharply with the size of the analyzed network.
The calculation of all EFMs of small networks (up to 50 reactions) is straightforward and simple.
However, despite the major progress made recently \cite{gagneur2004,terzer2008,jevremovic2011} the computation of the complete set of EFMs of large or even genome-scale networks is still impossible.
There is a number of tools specifically designed to calculate the complete set of EFMs as performant as possible,
such as {\it Metatool} \cite{kamp2006}, {\it CellNetAnalyzer} \cite{klamt2007} and {\it efmtool} \cite{terzer2008}.
To our best knowledge one of the fastest program currently available is {\it efmtool} by Marco Terzer which is written in the multi-platform programming language {\it Java},
supports multi-threading, is published under the open source software license {\it Simplified BSD Style License} \cite{osi}, and can be downloaded from \cite{efmtoolETH}. 

In the presented work we introduce a novel approach to speed up the computation of the complete set of biologically feasible EFMs by taking into account
the gene regulatory information of the investigated metabolic network.
Transcriptional regulatory networks (TRN) are typically provided as a boolean rule set, e.g. \cite{orth2009}.
These rules exclude many of the mathematically possible EFMs for biological reasons.
We implemented our algorithm by extending {\it efmtool}, thereby, exploiting the full power and advantage of open source software.
By utilizing a specific feature of the binary approach \cite{gagneur2004} which was applied in {\it efmtool},
the elimination of biologically infeasible modes can be done constantly and at an early stage of the EFM computation process.
Thereby, a significant reduction of the computational costs, such as execution time, memory consumption and harddisk space, is achieved.

\section{Methods}

\subsection{Binary approach}

Modern EFM computation programs, such as {\it efmtool}, use a binary approach \cite{gagneur2004} of the double description method \cite{fukuda1996}.
In the following we briefly review this binary approach.
We will introduce our modifications for the inclusion of transcriptional regulation in the next section.

The binary approach is characterized by splitting each mode into a binary part and a numerical part.
The binary part of a mode contains only a single bit for each reaction,
where '1' means that the reaction carries a flux and '0' stands for a reaction not carrying a flux.
While iterating through the binary algorithm, the numerical part of each mode is successively converted into the binary representation.
The iteration procedure terminates when each mode has been completely transformed to its binary form.
In a final post-iteration step the computed binary modes are converted back to their numerical forms.
The numerical representation of a mode gives the exact stoichiometric amount of each involved reaction that participates in the mode.

\begin{figure}
  \includegraphics[width=\columnwidth]{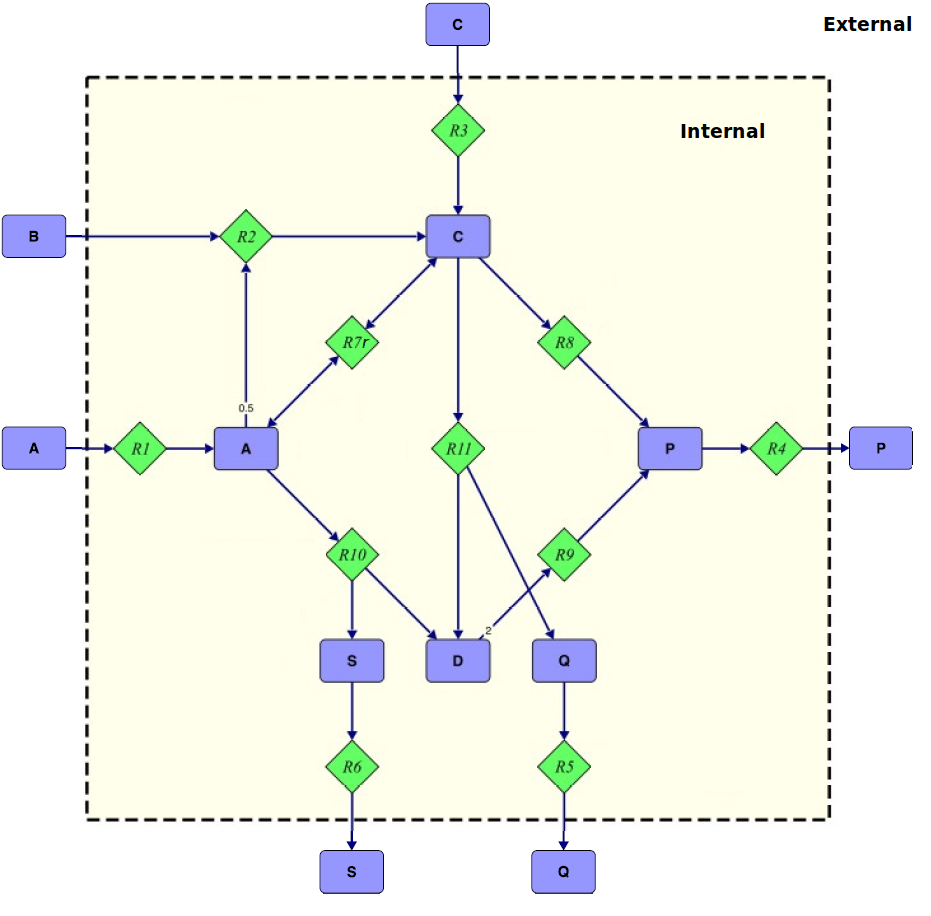}
  \caption{Example network consisting of 12 metabolites (rectangles) and 11 reactions (diamonds). Only one of the 11 reactions is reversible ({\it R7r}).\label{fig:exampleNetwork}}
\end{figure}

We demonstrate the general principals of the binary approach by the simple example network shown in Figure \ref{fig:exampleNetwork}.
For the sake of clarity the network will not be compressed in order to keep all originally specified reactions and metabolites of the network.
In a 'real-life' computation several compression strategies would be applied first in order to combine and remove topographically redundant reactions and metabolites \cite{gagneur2004}.
The example network consists of 11 reactions and 12 metabolites. Only reaction {\it R7r} is reversible.
The stoichiometric matrix S of the example network is shown in Table 1 of the supplementary data section.
The external metabolites do not obey the steady state condition and, thus, are irrelevant for the calculation of the EFMs.

First, the reversible reaction {\it R7r} is split into a forward and a backward irreversible reaction.
This is done by negating the column of the reversible reaction and appending the newly created column right after the original one.
Table \ref{tab:exampleNetworkExtendedStoichMatrix} shows the extended stoichiometric matrix S$_{ext}$ that only contains irreversible reactions.

\begin{table}
\centering
\resizebox{\columnwidth}{!}{
\begin{tabular}{l|rrrrrrrrrrrr}
     & R1 & R2   & R3 & R4 & R5 & R6 & R7f & R7b & R8 & R9 & R10 & R11 \\
\hline
A    & 1  & -0.5 & 0  &  0 &  0 &  0 & -1  &   1 &  0 &  0 & -1  &  0 \\
C    & 0  &  1   & 1  &  0 &  0 &  0 &  1  &  -1 & -1 &  0 &  0  & -1 \\
D    & 0  &  0   & 0  &  0 &  0 &  0 &  0  &   0 &  0 & -2 &  1  &  1 \\
P    & 0  &  0   & 0  & -1 &  0 &  0 &  0  &   0 &  1 &  1 &  0  &  0 \\
Q    & 0  &  0   & 0  &  0 & -1 &  0 &  0  &   0 &  0 &  0 &  0  &  1 \\
S    & 0  &  0   & 0  &  0 &  0 & -1 &  0  &   0 &  0 &  0 &  1  &  0 \\
\end{tabular}
}
\caption{Extended stoichiometric matrix S$_{ext}$ of the example network shown in Figure \ref{fig:exampleNetwork} after splitting the reversible reaction {\it R7r} into the two irreversible reactions {\it R7f} and {\it R7b}.}
\label{tab:exampleNetworkExtendedStoichMatrix}
\end{table}

The main process of computing all EFMs is based on the double description method \cite{fukuda1996}.
The basic principal of the double description method is to determine an initial set of solution modes
which are then iteratively combined and added to the set of existing modes until the complete set of modes is obtained.
The solution modes are stored in the mode matrix R that contains one column for each elementary mode.
Typically, the initialization of the mode matrix R is obtained by calculating the kernel K of the extended stoichiometric matrix S$_{ext}$.
The kernel K of the extended stoichiometric matrix S$_{ext}$ is defined by S$_{ext}$K = 0 
and is shown in Table \ref{tab:kernelMatrix}.

\begin{table}
\centering
\begin{tabular}{l|rrrrrr}
\hline
R1  & -0.5 &  0.5 & -0.5 & 0.5 & 1   & 0.5 \\
R2  & -1   & -1   &  1   & 1   & 0   & 1   \\
R3  & 1    &  0   &  0   & 0   & 0   & 0   \\
R4  & 0    &  0   &  0   & 1   & 0.5 & 0.5 \\ 
R5  & 0    &  0   &  0   & 0   & 0   & 1   \\
R6  & 0    &  0   &  0   & 0   & 1   & 0   \\
R7f & 0    &  1   &  0   & 0   & 0   & 0   \\
R7b & 0    &  0   &  1   & 0   & 0   & 0   \\
R8  & 0    &  0   &  0   & 1   & 0   & 0   \\
R9  & 0    &  0   &  0   & 0   & 0.5 & 0.5 \\
R10 & 0    &  0   &  0   & 0   & 1   & 0   \\
R11 & 0    &  0   &  0   & 0   & 0   & 1   \\
\end{tabular}
\caption{Kernel matrix K of the extended stoichiometric matrix shown in Table \ref{tab:exampleNetworkExtendedStoichMatrix} of the example network. }
\label{tab:kernelMatrix}
\end{table}

Next, the initial conversion to the binary representation of the mode matrix R is performed.
The final set of elementary modes of the extended network must only contain non-negative flux values,
as the extended network contains only irreversible reactions.
As pointed out by Gagneur and Klamt \cite{gagneur2004} using only irreversible reactions is of major importance,
as all non-zero elements of a mode will be retained if a new mode is created by combining this mode with other modes that have already been determined
during the calculation procedure.
Hence, all rows that contain only non-negative values can directly be transformed to the binary representation.
For the sake of clarity we use the character {\it f} for binary 1 indicating a {\it flux} carrying reaction
and the character {\it n} for binary 0 indicating that {\it no} flux occurs.
Usually, the initial solution matrix R is sorted in a way that all rows containing only positive values are at the top.
Table \ref{tab:initialModeMatrixR} shows the properly sorted mode matrix R containing numerical and binary values
before the iteration process is started.

\begin{table}
\centering
\begin{tabular}{l|cccccc}
    & M1 & M2 & M3 & M4 & M5 & M6 \\
\hline
R3  & f & n & n & n & n & n \\
R4  & n & n & n & f & f & f \\
R5  & n & n & n & n & n & f \\
R6  & n & n & n & n & f & n \\
R7f & n & f & n & n & n & n \\
R7b & n & n & f & n & n & n \\
R8  & n & n & n & f & n & n \\
R9  & n & n & n & n & f & f \\
R10 & n & n & n & n & f & n \\
R11 & n & n & n & n & n & f \\
R2  & -1 & -1 & 1 & 1 & 0 & 1 \\
R1  & -0.5 & 0.5 & -0.5 & 0.5 & 1 & 0.5 \\
\end{tabular}
\caption{Initial mode matrix R. The first ten rows are already transformed to the binary representation where {\it f} stands for binary 1 and indicates that the reaction carries a flux and {\it n} stands for binary 0 and indicates that no flux is carried. Note that the order of the reactions has changed compared to Table \ref{tab:exampleNetworkExtendedStoichMatrix} in order to maximize the number of rows that can be converted to binary form during the pre-iteration phase.}
\label{tab:initialModeMatrixR}
\end{table}

Next, the iteration procedure is performed.
Step by step each row that is still in numerical form is transformed to its binary representation.
As shown in Table \ref{tab:initialModeMatrixR} the next reaction to be processed is {\it R2}.
The double description method requires that all modes containing non-negative values at {\it R2} are retained,
whereas the modes with negative values are removed.
Furthermore, the method requires that all modes with negative values at {\it R2} are combined with adjacent modes exhibiting a positive value at {\it R2}.
Hence, the modes M1 and M2 are combined with M3, M4, and M6 resulting in six potential new modes.
Two modes are adjacent if the binary part of the new mode is not a superset of any already existing modes - except the two parent modes.
For the binary part, the combination of two adjacent modes is a simple and fast bitwise OR operation of the involved modes.
Combining the numerical part is achieved by a weighted subtraction of the two numerical vectors.
The new numerical value $v$ of row $r$ is calculated by $v_{new_{r}} = (v_{pos_{1}} v_{neg_{r}} - v_{neg_{1}} v_{pos_{r}})/(v_{pos_{1}} - v_{neg_{1}})$,
where $v_{{pos}_{r}}$ and $v_{{neg}_{r}}$ are the values of the positive and of the negative column at row $r$, respectively.
The row index $r$ runs from 1 to $n$, where row $r = 1$ is the row to be converted at current iteration step and
$n$ is the number of rows left to be converted.
Applying these instructions to the initial mode matrix R given in Table \ref{tab:initialModeMatrixR} results
in the new mode matrix shown in Table \ref{tab:step1ModeMatrixR}.

\begin{table}
\centering
\begin{tabular}{l|cccccccccc}
    & M1 & M2 & M3 & M4 & M5 & M6 & M7 & M8 & M9 & M10 \\
\hline
R3  &     n &   n & n &   n &   n &   n &   n &   f &   f &    f \\
R4  &     n &   f & f &   f &   f &   f &   n &   f &   f &    n \\
R5  &     n &   n & n &   f &   f &   n &   n &   f &   n &    n \\
R6  &     n &   n & f &   n &   n &   n &   n &   n &   n &    n \\
R7f &     n &   n & n &   n &   f &   f &   f &   n &   n &    n \\
R7b &     f &   n & n &   n &   n &   n &   f &   n &   n &    f \\
R8  &     n &   f & n &   n &   n &   f &   n &   n &   f &    n \\
R9  &     n &   n & f &   f &   f &   n &   n &   f &   n &    n \\
R10 &     n &   n & f &   n &   n &   n &   n &   n &   n &    n \\
R11 &     n &   n & n &   f &   f &   n &   n &   f &   n &    n \\
R2  &     f &   f & n &   f &   n &   n &   n &   n &   n &    n \\
R1  &  -0.5 & 0.5 & 1 & 0.5 & 0.5 & 0.5 & 0.0 & 0.0 & 0.0 & -0.5 \\
\end{tabular}
\caption{Mode matrix R after the first iteration step converting reaction {\it R2} from numerical to binary form.}
\label{tab:step1ModeMatrixR}
\end{table}

Applying the mode combination procedure again for the last row to be converted ({\it R1}) results in the final mode matrix R as shown in Table \ref{tab:step2ModeMatrixR}.
Now, the matrix R contains only binary elements.
Note that the performance of the described iteration procedure for 'real-life' networks can tremendously be increased
if tree structures are utilized to store the binary representation of the modes \cite{terzer2008}.

\begin{table}
\centering
\resizebox{\columnwidth}{!}{
\begin{tabular}{l|cccccccccccc}
    & M1 & M2 & M3 & M4 & M5 & M6 & M7 & M8 & M9 & M10 & M11 & M12 \\
\hline
R3  &   n & n &   n &   n &   n &   n &   f &   f &  f &  n & n & n \\
R4  &   f & f &   f &   f &   f &   n &   f &   f &  f &  f & f & f \\
R5  &   n & n &   f &   f &   n &   n &   f &   n &  n &  f & n & n \\
R6  &   n & f &   n &   n &   n &   n &   n &   n &  f &  n & f & n \\
R7f &   n & n &   n &   f &   f &   f &   n &   n &  n &  n & n & n \\
R7b &   n & n &   n &   n &   n &   f &   n &   n &  f &  f & f & f \\
R8  &   f & n &   n &   n &   f &   n &   n &   f &  n &  n & n & f \\
R9  &   n & f &   f &   f &   n &   n &   f &   n &  f &  f & f & n \\
R10 &   n & f &   n &   n &   n &   n &   n &   n &  f &  n & f & n \\
R11 &   n & n &   f &   f &   n &   n &   f &   n &  n &  f & n & n \\
R2  &   f & n &   f &   n &   n &   n &   n &   n &  n &  f & f & f \\
R1  &   f & f &   f &   f &   f &   n &   n &   n &  n &  n & n & n \\
\end{tabular}
}
\caption{Mode matrix R after the final iteration step containing only binary values. Mode M6 is the futile 2-cycle mode and can be removed.}
\label{tab:step2ModeMatrixR}
\end{table}

Next, the futile 2-cycle mode (M6) that was created by splitting the reversible reaction {\it R7r} is removed.
Then the irreversible forward and backward reactions {\it R7f} and {\it R7b} are combined by a simple bitwise OR operation
in order to obtain the reversible reaction {\it R7r} again.
The final set of modes in binary form is shown in Table \ref{tab:binaryEFM}.
Reconverting the binary form to the numerical representation is achieved by using the fact
that the reduced null space matrix $N_{red}$ multiplied by the sought numerical mode has dimension 1 and is equal to zero.
$N_{red}$ is a sub-matrix of the kernel K that only contains columns/reactions where the binary mode to be transformed carries a flux.
Hence, only a homogeneous linear system has to be solved to obtain the 1-dimensional solution vector that represents the numerical form of a mode.
The result of the reconversion of the binary modes for the simple example network is listed in Table \ref{tab:numericEFM}.

\begin{table}
\centering
\resizebox{\columnwidth}{!}{
\begin{tabular}{l|rrrrrrrrrrr}
     & R1 & R2 & R3 & R4 & R5 & R6 & R7r & R8 & R9 & R10 & R11 \\
\hline
EFM01 & 0 & 1 & 0 & 1 & 0 & 0 & 1 & 1 & 0 & 0 & 0 \\
EFM02 & 0 & 1 & 0 & 1 & 1 & 0 & 1 & 0 & 1 & 0 & 1 \\
EFM03 & 0 & 1 & 0 & 1 & 0 & 1 & 1 & 0 & 1 & 1 & 0 \\
EFM04 & 0 & 0 & 1 & 1 & 0 & 1 & 1 & 0 & 0 & 1 & 0 \\
EFM05 & 1 & 1 & 0 & 1 & 0 & 0 & 0 & 1 & 0 & 0 & 0 \\
EFM06 & 1 & 1 & 0 & 1 & 1 & 0 & 0 & 0 & 1 & 0 & 1 \\
EFM07 & 1 & 0 & 0 & 1 & 0 & 1 & 0 & 0 & 1 & 1 & 0 \\
EFM08 & 1 & 0 & 0 & 1 & 1 & 0 & 1 & 0 & 1 & 0 & 1 \\
EFM09 & 1 & 0 & 0 & 1 & 0 & 0 & 1 & 1 & 0 & 0 & 0 \\
EFM10 & 0 & 0 & 1 & 1 & 1 & 0 & 0 & 0 & 1 & 0 & 1 \\
EFM11 & 0 & 0 & 1 & 1 & 0 & 0 & 0 & 1 & 0 & 0 & 0 \\
\end{tabular}
}
\caption{Binary representation of all elementary flux modes of the example network shown in Figure \ref{fig:exampleNetwork}. 1 means that the reaction carries a flux and 0 means the reaction carries no flux. Note that the futile two-cycle of the reversible reaction {\it R7r} has already been removed and the forward and backward irreversible reactions ({\it R7f} and {\it Rfb}) have been combined to the reversible reaction {\it R7r} by a bitwise OR operation.}
\label{tab:binaryEFM}
\end{table}

\begin{table}
\centering
\resizebox{\columnwidth}{!}{
\begin{tabular}{l|rrrrrrrrrrr}
     & R1 & R2   & R3 & R4 & R5 & R6 & R7r & R8 & R9 & R10 & R11 \\
\hline
EFM01 & 0.0 & 1.0 & 0.0 & 0.5  & 0.0 & 0.0 & -0.5 & 0.5 & 0.0  & 0.0 & 0.0 \\
EFM02 & 0.0 & 1.0 & 0.0 & 0.25 & 0.5 & 0.0 & -0.5 & 0.0 & 0.25 & 0.0 & 0.5 \\
EFM03 & 0.0 & 1.0 & 0.0 & 0.25 & 0.0 & 0.5 & -1.0 & 0.0 & 0.25 & 0.5 & 0.0 \\
EFM04 & 0.0 & 0.0 & 1.0 & 0.5  & 0.0 & 1.0 & -1.0 & 0.0 & 0.5  & 1.0 & 0.0 \\
EFM05 & 0.5 & 1.0 & 0.0 & 1.0  & 0.0 & 0.0 &  0.0 & 1.0 & 0.0  & 0.0 & 0.0 \\
EFM06 & 0.5 & 1.0 & 0.0 & 0.5  & 1.0 & 0.0 &  0.0 & 0.0 & 0.5  & 0.0 & 1.0 \\
EFM07 & 1.0 & 0.0 & 0.0 & 0.5  & 0.0 & 1.0 &  0.0 & 0.0 & 0.5  & 1.0 & 0.0 \\
EFM08 & 1.0 & 0.0 & 0.0 & 0.5  & 1.0 & 0.0 &  1.0 & 0.0 & 0.5  & 0.0 & 1.0 \\
EFM09 & 1.0 & 0.0 & 0.0 & 1.0  & 0.0 & 0.0 &  1.0 & 1.0 & 0.0  & 0.0 & 0.0 \\
EFM10 & 0.0 & 0.0 & 1.0 & 0.5  & 1.0 & 0.0 &  0.0 & 0.0 & 0.5  & 0.0 & 1.0 \\
EFM11 & 0.0 & 0.0 & 1.0 & 1.0  & 0.0 & 0.0 &  0.0 & 1.0 & 0.0  & 0.0 & 0.0 \\
\end{tabular}
}
\caption{Numerical representation of all EFM of the example network shown in Figure \ref{fig:exampleNetwork}.}
\label{tab:numericEFM}
\end{table}

The binary approach combines several essential advantages:
a) modes are stored in binary format which dramatically reduces the memory usage,
b) new modes are calculated from existing adjacent modes by using simple bitwise boolean functions which are very fast compared to numeric operations, and
c) the bitwise boolean operations used are 'exact', hence, numerical accuracy problems are minimized.

\subsection{Gene regulation information}

TRN control the process of gene expression in cells.
They determine how genes activate or repress certain fluxes.
Hence, the gene regulatory information of a network imposes additional constraints on the reactions,
and, as a consequence, reduces the solution space resulting in a lower number of biologically feasible EFMs.
Typically, the gene regulation information is provided in form of boolean functions \cite{orth2009},
such as the NOT, OR, and AND operations.

\begin{figure}
  \includegraphics[width=\columnwidth]{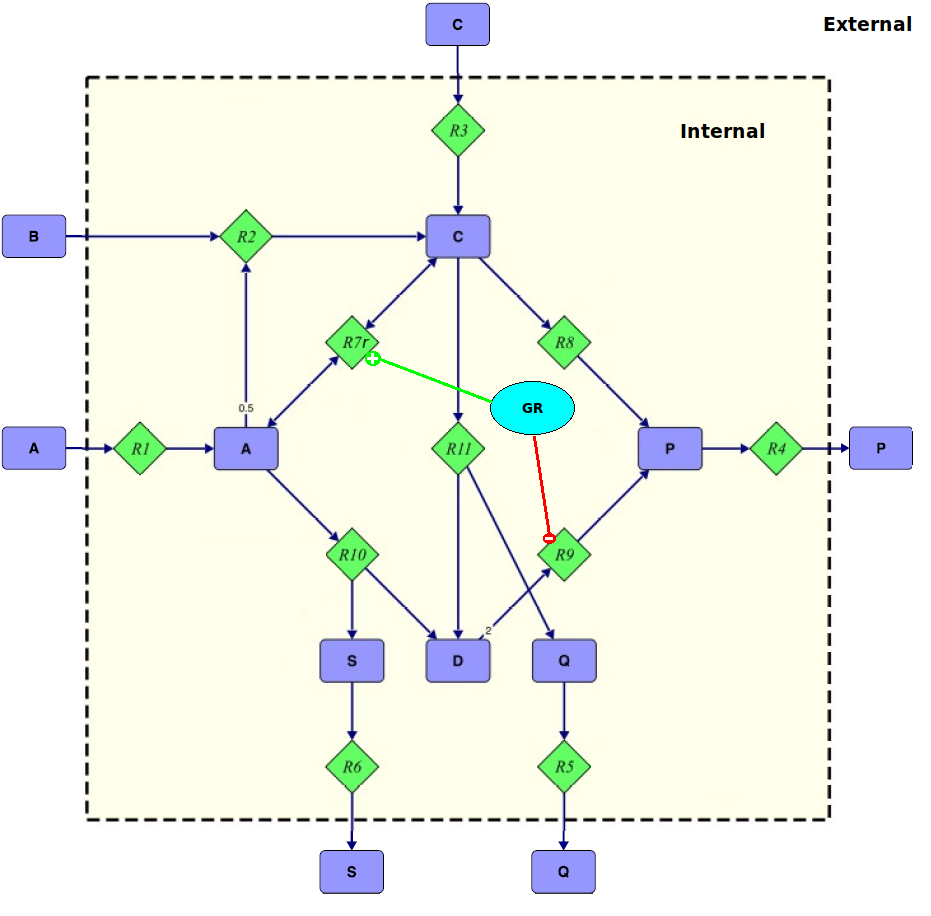}
  \caption{Example network including the gene regulator network: {\it R7r} = NOT({\it R9}).\label{fig:exampleNetworkGeneReg}}
\end{figure}

As illustrated in Figure \ref{fig:exampleNetworkGeneReg} we assume
that the gene regulatory network of the example network shown in Figure \ref{fig:exampleNetwork}
only consists of a gene {\it GR} that activates reaction {\it R7r} and deactivates reactions {\it R9}.
The function of gene {\it GR} can be transformed to a single boolean expression: {\it R7r} = NOT({\it R9}).
This constraint means that the reaction {\it R7r} must not carry a flux
when reaction {\it R9} carries a flux and vice versa.
A simple approach to get the reduced solution space
is the application of this gene regulatory rule after
all mathematically possible modes were calculated.
Naturally, this method does not result in any performance improvement.
However, if we consider the basic principle of the binary approach
described above, a significant speed up of the computation process can be obtained.
The boolean operation {\it R7r} = NOT({\it R9}) implies that
the rule is not obeyed if:
a) {\it R9} = 1 = {\it f} and {\it R7r} = 1 = {\it f} or
b) {\it R9} = 0 = {\it n} and {\it R7r} = 0 = {\it n}.
The first expression is of particular interest,
as it can be used to eliminate all modes from the solution matrix R - at any time of the iteration process -
if {\it R9} and {\it R7r} do carry a flux.
This statement is true, as
a) the considered mode itself disobeys the rule and
b) all children modes generated from the considered mode by combination with other modes will also disobey the rule.
The latter property is owed to the fact that a flux value at a certain reaction will be retained by the bitwise OR operation
for the rest of the computation procedure (see previous subsection for further details).
Removing a mode as soon as possible is of essential importance,
as this mode is hindered to create offspring modes that would have to be eliminated at a later stage.
The second expression (if {\it R9} = 0 = {\it n} and {\it R7r} = 0 = {\it n}) is of no use during the iteration process,
as a {\it no} flux value of {\it R9} or {\it R7r} can become a flux carrying reaction in a child mode that
is created in a later iteration step.
Hence, removing a currently disobeying mode with {\it R9} = 0 and {\it R7r} = 0 would result in the loss of
children modes that obey the rule {\it R7r} = NOT({\it R9}).
However, the rule {\it R7r} = NOT({\it R9}) for {\it R9} = 0 and {\it R7r} = 0 can still be used
to remove infeasible modes after finishing the computation of all binary modes

The above considerations make clear that there are two types of rules:
a) rules that can be applied during the iteration process and
b) rules that can be applied during the post-processing step after finishing the mode calculation.

Determining if a boolean rule {\it Ro} = $\mathcal{B}$({\it R1},...,{\it Rn}) qualifies for the iteration phase is simple.
If the output reaction {\it Ro} of the rule is 0 (does not carry a flux) when all input reactions {\it R1},...,{\it Rn} are 1 (carry a flux)
then the rule can be used during the iteration phase.

Special care must be taken for reversible reactions, as they are split and, hence, occur twice in the extended set of reactions.
If either the forward or the backward reaction carries a flux then the original reaction is supposed to be flux carrying
when checked against a boolean rule.

Applying these concepts to the example network with the gene regulatory rule {\it R7r} = NOT({\it R9}) results
in a mode matrix R after the first iteration step as shown in Table \ref{tab:step1ModeMatrixRApplyRules}.
Table \ref{tab:step1ModeMatrixRApplyRules} highlights in red font color all reactions involved in rule {\it R7r} = NOT({\it R9}) that carry a flux.
It can be seen that mode M5 disobeys the rule and is removed from the matrix R.

\begin{table}
\centering
\resizebox{\columnwidth}{!}{
\begin{tabular}{l|cccc>{\columncolor{dg}}cccccc}
    &         M1 &        M2 & M3 &        M4 & M5 & M6 & M7 & M8 & M9 & M10 \\
\hline
R3  &          n &   n &         n &         n &         n &         n &         n &         f &   f &         f \\
R4  &          n &   f &         f &         f &         f &         f &         n &         f &   f &         n \\
R5  &          n &   n &         n &         f &         f &         n &         n &         f &   n &         n \\
R6  &          n &   n &         f &         n &         n &         n &         n &         n &   n &         n \\
R7f &          n &   n &         n &         n & \tc{r}{f} & \tc{r}{f} & \tc{r}{f} &         n &   n &         n \\
R7b &  \tc{r}{f} &   n &         n &         n &         n &         n & \tc{r}{f} &         n &   n & \tc{r}{f} \\
R8  &          n &   f &         n &         n &         n &         f &         n &         n &   f &         n \\
R9  &          n &   n & \tc{r}{f} & \tc{r}{f} & \tc{r}{f} &         n &         n & \tc{r}{f} &   n &         n \\
R10 &          n &   n &         f &         n &         n &         n &         n &         n &   n &         n \\
R11 &          n &   n &         n &         f &         f &         n &         n &         f &   n &         n \\
R2  &          f &   f &         n &         f &         n &         n &         n &         n &   n &         n \\
R1  &       -0.5 & 0.5 &         1 &       0.5 &       0.5 &       0.5 &       0.0 &       0.0 & 0.0 &      -0.5 \\
\end{tabular}
}
\caption{Mode matrix R after the first iteration step. The red font color highlights reactions involved in rule {\it R7r} = NOT({\it R9}) that carry a flux. Mode M5 (highlighted in grey background color) disobeys the rule and is removed from the matrix.}
\label{tab:step1ModeMatrixRApplyRules}
\end{table}

In the next step mode M5 does not exist and, hence, fewer adjacency tests have to be performed.
Table \ref{tab:step2ModeMatrixRApplyRules} shows the mode matrix R after the final iteration step.
It can be seen that mode M8, M9, and M10 do not obey the iteration phase rule,
as {\it R9} and {\it R7f} or {\it R7b} carry a flux (highlighted by the red font color).
Hence, M8, M9, and M10 can be removed during the iteration phase.

\begin{table}
\centering
\resizebox{\columnwidth}{!}{
\begin{tabular}{l|>{\columncolor{lg}}cccccc>{\columncolor{lg}}c>{\columncolor{dg}}c>{\columncolor{dg}}c>{\columncolor{dg}}cc}
    & M1 & M2 & M3 & M4 & M5 & M6 & M7 & M8 & M9 & M10 & M11 \\
\hline
R3  &         n &         n &         n &         n &         n &         f &         f &         f &         n &        n &        n \\
R4  &         f &         f &         f &         f &         n &         f &         f &         f &         f &        f &        f \\
R5  &         n &         n &         f &         n &         n &         f &         n &         n &         f &        n &        n \\
R6  &         n &         f &         n &         n &         n &         n &         n &         f &         n &        f &        n \\
R7f & \tc{b}{n} & \tc{b}{n} & \tc{b}{n} & \tc{r}{f} & \tc{r}{f} & \tc{b}{n} & \tc{b}{n} &         n &         n &        n &        n \\
R7b & \tc{b}{n} & \tc{b}{n} & \tc{b}{n} &         n & \tc{r}{f} & \tc{b}{n} & \tc{b}{n} & \tc{r}{f} & \tc{r}{f} & \tc{r}{f} & \tc{r}{f} \\
R8  &         f &         n &         n &         f &         n &         n &         f &         n &         n &        n &        f \\
R9  & \tc{b}{n} & \tc{r}{f} & \tc{r}{f} & \tc{b}{n} & \tc{b}{n} & \tc{r}{f} & \tc{b}{n} & \tc{r}{f} & \tc{r}{f} & \tc{r}{f} & \tc{b}{n} \\
R10 &         n &         f &         n &         n &         n &         n &         n &         f &         n &        f &        n \\
R11 &         n &         n &         f &         n &         n &         f &         n &         n &         f &        n &        n \\
R2  &         f &         n &         f &         n &         n &         n &         n &         n &         f &        f &        f \\
R1  &         f &         f &         f &         f &         n &         n &         n &         n &         n &        n &        n \\
\end{tabular}
}
\caption{Mode matrix R after the final iteration step. Mode M8, M9, and M10 do not obey the iteration phase rule. Additionally, mode M1 and M7 disobey the post-processing rule. M5 is also removed, as it is the futile 2-cycle mode created by splitting the reversible reaction {\it R7r} into two irreversible reaction.}
\label{tab:step2ModeMatrixRApplyRules}
\end{table}

Furthermore, Table \ref{tab:step2ModeMatrixRApplyRules} illustrates that mode M1 and M7
disobey the post-processing rule, as {\it R9}, {\it R7f}, and {\it R7b} do not carry a flux value.
Consequently, after removing the futile 2-cycle mode M5 the final mode matrix R only contains the five modes M2, M3, M4, M6, and M11.
Before transforming the binary modes back to their numerical form the split irreversible reactions {\it R7f} and {\it R7b}
must be combined to the reversible reaction {\it R7r}.
The final set of feasible EFMs is listed in Table \ref{tab:numericEFMApplyRules}.

\begin{table}
\centering
\resizebox{\columnwidth}{!}{
\begin{tabular}{l|rrrrrrrrrrr}
     & R1 & R2   & R3 & R4 & R5 & R6 & R7r & R8 & R9 & R10 & R11 \\
\hline
EFM01 & 0.5 & 1.0 & 0.0 & 0.5  & 1.0 & 0.0 &  0.0 & 0.0 & 0.5  & 0.0 & 1.0 \\
EFM02 & 0.0 & 1.0 & 0.0 & 0.5  & 0.0 & 0.0 & -0.5 & 0.5 & 0.0  & 0.0 & 0.0 \\
EFM03 & 1.0 & 0.0 & 0.0 & 0.5  & 0.0 & 1.0 &  0.0 & 0.0 & 0.5  & 1.0 & 0.0 \\
EFM04 & 1.0 & 0.0 & 0.0 & 1.0  & 0.0 & 0.0 &  1.0 & 1.0 & 0.0  & 0.0 & 0.0 \\
EFM05 & 0.0 & 0.0 & 1.0 & 0.5  & 1.0 & 0.0 &  0.0 & 0.0 & 0.5  & 0.0 & 1.0 \\
\end{tabular}
}
\caption{Numerical representation of all EFMs of the example network shown in Figure \ref{fig:exampleNetwork} if the gene regulatory rule {\it R7r} = NOT({\it R9}) is applied.}
\label{tab:numericEFMApplyRules}
\end{table}

\subsection{Implementation}

We implemented our approach as an extension to the open source software {\it efmtool}.
The mode elimination algorithm was realized by adding three {\it Java} packages to the original version of {\it efmtool}.
The three packages contained ten new {\it Java} classes.
These new classes are responsible for handling the genetic rules and checking the modes against them.
Two already existing {\it Java} classes were slightly enhanced in order to invoke the mode check.
The boolean rules are provided by an additional input file using the command line argument {\tt -generule}.
The extended version of {\it efmtool} was compiled by {\it JDK 1.7.0}.
The implementation of the extension was performed as non-invasive as possible,
which means that the performance gain might be even better if the new method is integrated to {\it efmtool} more thoroughly.
The mode checks for the iteration phase were implemented using binary bit patterns
where the patterns are created simply by setting the involved reactions (all input reactions and the output reaction) to 1
\footnote{Note that {\it efmtool} uses an inverse logic where 0 stands for flux carrying reactions and 1 stands for not carrying a flux. Hence, the involved bitwise operations and comparison have to be changed accordingly.}.
If a tested mode has every bits set that occurs in the binary bit pattern of a rule then the mode is eliminated.
The mode check for the post-processing step was realized by utilizing a reverse polish notation approach
that allows a simple and fast execution of boolean functions with any values for the input reactions.
The general sequence of operation of our extended version of the binary double description method is shown in the supplementary data section.

\section{Results and Discussion}

We tested our approach on the E. coli core model provided by \cite{orth2009,EcoliCoreUCSanDiego}.
The model consists of 94 metabolites and 95 reactions. 59 reactions are reversible.
Gene regulatory information for this model is provided by \cite{EcoliCoreUCSanDiego}
in form of a gene-enzyme-reaction mapping.
The mapping was checked for consistency and contradicting rules were removed from the provided data set.
The final mapping used by our algorithm to exclude infeasible EFMs contained 58 boolean functions and is shown in Table 2 of the supplementary information section.
Only four of these 58 rules qualify for being used during the iteration phase of the EFM computation procedure
and are listed in Table \ref{tab:finalMapping}.

\begin{table}
\centering
\begin{tabular}{lcl}
\toprule
Booleanly combined & & Required value of \\
input reactions    & & effected output reaction \\
\toprule

R\_EX\_glc\_e = 1 & $\longrightarrow$ & R\_EX2\_ac\_e = 0 \\
\hline
R\_EX\_glc\_e = 1 & $\longrightarrow$ & R\_ICL = 0 \\
\hline
R\_EX\_glu\_L\_e = 1 & $\longrightarrow$ & R\_GLUDy = 0 \\
\hline
R\_EX\_glu\_L\_e = 1 & $\longrightarrow$ & R\_GLUSy = 0 \\
\hline
\end{tabular}
\caption{ The four boolean rules used by the introduced elimination algorithm to exclude biologically infeasible EFMs during the iteration phase.
The 54 other rules can only be applied during the post-processing phase and are shown in the supplementary data section.}
\label{tab:finalMapping}
\end{table}

Table \ref{tab:compEcoliRuns} compares a regular run without regulatory information and a run using the available gene regulation rules.
Both runs were performed on a Linux Ubuntu 11.04 computer with 2 Intel Xeon CPUs (6 cores each) and a total of 192 GB of RAM using 10 parallel threads.
The table shows that after the iteration phase 226 million modes were obtained without regulatory information.
Despite the fact that just four of the 58 boolean rules qualify for the iteration phase,
applying the novel mode elimination approach resulted in only 76.7 million modes.
This mode removal during the iteration phase caused a reduction of the iteration runtime from 30.9 to 5.3 hours
and a decrease of the maximum number of adjacent candidates from 2.2$\cdot$10$^{15}$ to 2.6$\cdot$10$^{14}$.
The elimination of modes in the post-processing phase was even more pronounced,
as only 2.1 million of the 76.7 million modes adhered to the boolean gene-enzyme-reaction mapping rules.
Interestingly, even the post-processing runtime is reduced by our approach (from 3.2 to 1.8 hours)
although 76.7 million binary modes were applied to the post-processing boolean rules.
This behaviour is owed to the fact that only 2.6 million modes survive this post-processing filtering procedure
which means that only a fraction of the modes were converted from the binary to the numerical representation
and were written to disk.
This huge reduction of the total number of modes (226.3 million to 2.1 million) had a major influence
on the required harddisk space which was decreased from 251 GB to 2.3 GB if an uncompressed double precision text format was used.
Table \ref{tab:compEcoliRuns} clearly shows that considering gene regulatory information in the computation process
has a huge impact on the computational key properties of the calculation of EFMs.

In order to verify that the the presented extension of the {\it efmtool} computes the correct EFMs,
an extra software tool was developed that applies the boolean rules to the complete and unfiltered set of EFMs.
The two tools computed identical sets of EFMs ensuring that the {\it efmtool} extension produces the correct result.

\begin{table}
\centering
\resizebox{\columnwidth}{!}{
\begin{tabular}{l|r|r}
                               & w/o gene regulation  & with gene regulation \\
\hline
No. of modes (iteration)       & $226.3 \cdot 10^{6}$ & $76.7 \cdot 10^{6}$ \\
No. of modes (post-processing) & $226.3 \cdot 10^{6}$ & $2.1\cdot 10^{6}$  \\
Max. adjacent candidates       & $2.2 \cdot 10^{15}$  & $2.6 \cdot 10^{14}$ \\
Max. RAM usage                 & 153 GB               & 73 GB               \\
Runtime (iteration)            & 30.9 h               & 5.3 h               \\
Runtime (post-processing)      &  3.2 h               & 1.8 h               \\
Runtime (total)                & 34.1 h               & 7.1 h               \\
Disk space                     & 251 GB               & 2.3 GB              \\
\end{tabular}
}
\caption{Comparison of EFM calculation with and without taking into account gene regulatory information.
The required disk space is given for a result file containing all modes in text format using double precision.
The iteration runtime is the time spent creating the binary modes without pre- and post-processing
and the line 'max. adjacent candidates' shows the maximum number of potentially occurring adjacent pairs. }
\label{tab:compEcoliRuns}
\end{table}

Table \ref{tab:compNumModes} shows the development of the number of obtained modes as a function of finished iterations.
In total, 21 iteration steps were performed in order to compute the complete set of EFMs.
Up to iteration 9 not a single EFM was eliminated and the inclusion of gene regulatory information had no effect.
The first removal occurred at iteration 10, where 3 modes were deleted.
Although in total only 1.6 million modes were removed during the iteration phase,
the final number of modes was reduced by 149.6 million modes.
This huge reduction is a result of lost parent modes which otherwise could have spawned a multitude of new children.

\begin{table}
\centering
\resizebox{\columnwidth}{!}{
\begin{tabular}{l|r|r|r}
Iteration & No. of removed   & No. of modes incl. & No. of modes w/o \\
 No.      & infeasible modes & gene regulation    & gene regulation   \\
\hline
9   &         0        &            97         &              97           \\
10  &         3        &            205        &              208          \\
11  &         0        &            285        &              288          \\
12  &         0        &            454        &              457          \\
13  &         0        &            456        &              459          \\
14  &         0        &            751        &              755          \\
15  &         0        &            849        &              952          \\
16  &         604      &            2,113      &              3,223        \\
17  &         3        &            6,463      &              9,454        \\
18  &         850      &            17,154     &              28,114       \\
19  &         2,168    &            27,468     &              48,388       \\
20  &         0        &            27,468     &              48,388       \\
21  &         1,597    &            57,180     &              112,180      \\
22  &         1,717    &            244,858    &              486,847      \\
23  &         81       &            224,537    &              444,371      \\
24  &         93,933   &            519,853    &              1,243,347    \\
25  &         109,042  &            1,701,029  &              4,566,570    \\
26  &         295,410  &            2,832,654  &              8,012,612    \\
27  &         31,045   &            4,505,295  &              12,790,524   \\
28  &         250,704  &            12,895,654 &              37,465,244   \\
29  &         247,738  &            26,365,168 &              77,934,795   \\
30  &         59,107   &            32,421,087 &              94,929,161   \\
31  &         591,718  &            76,690,502 &              226,269,046  \\
\hline
sum &        1,685,720 &                       &                           \\
\end{tabular}
}
\caption{Comparison of the number of EFMs as a function of the iteration step for simulations
with gene regulatory information and without gene regulatory information. }
\label{tab:compNumModes}
\end{table}

In order to find an initial value of the mode matrix R the kernel of the extended stoichiometric matrix is computed.
Before the iteration phase is started the reactions are sorted.
This is done to put all reactions with only positive values to the top
which results in the maximum number of reactions that can be transformed to the binary form
before the iteration procedure is even started.
Several approaches can be applied to sort the reactions that also contain negative values,
e.g. taking no special measures (random order) or ordering by increasing potential combinations
(number of negative values times number of positive values).
As the iteration phase rules can only be applied if the involved reactions are already converted to the binary representation,
it seems beneficial to convert all reactions that are involved in iteration phase rules to the binary form as soon as possible.
This concept requires that all these reactions are moved to the top of the set of rows containing negative values.
Note that this approach was not implemented in the developed extension but could result in an additional performance gain if realized.

\section{Conclusion}

We implemented a novel approach to speed up the computation of the complete set of EFMs of a metabolic network
by extending the open source program {\it efmtool} written by Marco Terzer.
Our extension allows the consideration of gene-enzyme-reaction mappings in the process of EFM computation.
Biologically infeasible flux modes are constantly eliminated during the calculation process.
By implementing an early stage exclusion of modes a considerable reduction of computational costs was achieved
which pushes the maximum size of calculable networks to new limits.
We think that our approach is another step to the final goal of studying genome-scale metabolic networks by elementary flux modes.

\section{Acknowledgments}

The authors gratefully acknowledge the support by the Federal Ministry of Economy, Family and Youth (BMWFJ), the Federal Ministry of Traffic, Innovation and Technology (bmvit), the Styrian Business Promotion Agency SFG, the Standortagentur Tirol and ZIT - Technology Agency of the City of Vienna through the COMET-Funding Program managed by the Austrian Research Promotion Agency FFG.

\bibliography{efmtool_extension}

\begin{thebibliography}{14}%
\makeatletter
\providecommand \@ifxundefined [1]{%
 \@ifx{#1\undefined}
}%
\providecommand \@ifnum [1]{%
 \ifnum #1\expandafter \@firstoftwo
 \else \expandafter \@secondoftwo
 \fi
}%
\providecommand \@ifx [1]{%
 \ifx #1\expandafter \@firstoftwo
 \else \expandafter \@secondoftwo
 \fi
}%
\providecommand \natexlab [1]{#1}%
\providecommand \enquote  [1]{``#1''}%
\providecommand \bibnamefont  [1]{#1}%
\providecommand \bibfnamefont [1]{#1}%
\providecommand \citenamefont [1]{#1}%
\providecommand \href@noop [0]{\@secondoftwo}%
\providecommand \href [0]{\begingroup \@sanitize@url \@href}%
\providecommand \@href[1]{\@@startlink{#1}\@@href}%
\providecommand \@@href[1]{\endgroup#1\@@endlink}%
\providecommand \@sanitize@url [0]{\catcode `\\12\catcode `\$12\catcode
  `\&12\catcode `\#12\catcode `\^12\catcode `\_12\catcode `\%12\relax}%
\providecommand \@@startlink[1]{}%
\providecommand \@@endlink[0]{}%
\providecommand \url  [0]{\begingroup\@sanitize@url \@url }%
\providecommand \@url [1]{\endgroup\@href {#1}{\urlprefix }}%
\providecommand \urlprefix  [0]{URL }%
\providecommand \Eprint [0]{\href }%
\@ifxundefined \urlstyle {%
  \providecommand \doi  [0]{\begingroup \@sanitize@url \@doi}%
  \providecommand \@doi [1]{\endgroup \@@startlink {\doibase
  #1}doi:\discretionary {}{}{}#1\@@endlink }%
}{%
  \providecommand \doi  [0]{doi:\discretionary{}{}{}\begingroup
  \urlstyle{rm}\Url }%
}%
\providecommand \doibase [0]{http://dx.doi.org/}%
\providecommand \Doi [0]{\begingroup \@sanitize@url \@Doi }%
\providecommand \@Doi  [1]{\endgroup\@@startlink{\doibase#1}\@@Doi}%
\providecommand \@@Doi [1]{#1\@@endlink}%
\providecommand \selectlanguage [0]{\@gobble}%
\providecommand \bibinfo  [0]{\@secondoftwo}%
\providecommand \bibfield  [0]{\@secondoftwo}%
\providecommand \translation [1]{[#1]}%
\providecommand \BibitemOpen [0]{}%
\providecommand \bibitemStop [0]{}%
\providecommand \bibitemNoStop [0]{.\EOS\space}%
\providecommand \EOS [0]{\spacefactor3000\relax}%
\providecommand \BibitemShut  [1]{\csname bibitem#1\endcsname}%
\bibitem [{\citenamefont {Schuster}\ \emph {et~al.}(2000)\citenamefont
  {Schuster}, \citenamefont {Fell},\ and\ \citenamefont
  {Dandekar}}]{schuster2000}%
  \BibitemOpen
  \bibfield  {author} {\bibinfo {author} {\bibfnamefont {S.}~\bibnamefont
  {Schuster}}, \bibinfo {author} {\bibfnamefont {D.~A.}\ \bibnamefont {Fell}},
  \ and\ \bibinfo {author} {\bibfnamefont {T.}~\bibnamefont {Dandekar}},\ }\Doi
  {10.1038/73786} {\bibfield  {journal} {\bibinfo  {journal} {Nat Biotech},\
  }\textbf {\bibinfo {volume} {18}},\ \bibinfo {pages} {326} (\bibinfo {year}
  {2000})},\ ISSN \bibinfo {issn} {1087-0156}\BibitemShut {NoStop}%
\bibitem [{\citenamefont {Schuster}\ \emph {et~al.}(1999)\citenamefont
  {Schuster}, \citenamefont {Dandekar},\ and\ \citenamefont
  {Fell}}]{schuster1999}%
  \BibitemOpen
  \bibfield  {author} {\bibinfo {author} {\bibfnamefont {S.}~\bibnamefont
  {Schuster}}, \bibinfo {author} {\bibfnamefont {T.}~\bibnamefont {Dandekar}},
  \ and\ \bibinfo {author} {\bibfnamefont {D.~A.}\ \bibnamefont {Fell}},\ }\Doi
  {10.1016/S0167-7799(98)01290-6} {\bibfield  {journal} {\bibinfo  {journal}
  {Trends in Biotechnology},\ }\textbf {\bibinfo {volume} {17}},\ \bibinfo
  {pages} {53} (\bibinfo {year} {1999})},\ ISSN \bibinfo {issn}
  {0167-7799}\BibitemShut {NoStop}%
\bibitem [{\citenamefont {Klamt}\ and\ \citenamefont
  {Stelling}(2002)}]{klamt2002}%
  \BibitemOpen
  \bibfield  {author} {\bibinfo {author} {\bibfnamefont {S.}~\bibnamefont
  {Klamt}}\ and\ \bibinfo {author} {\bibfnamefont {J.}~\bibnamefont
  {Stelling}},\ }\Doi {10.1023/A:1020390132244} {\bibfield  {journal} {\bibinfo
   {journal} {Molecular Biology Reports},\ }\textbf {\bibinfo {volume} {29}},\
  \bibinfo {pages} {233} (\bibinfo {year} {2002})},\ ISSN \bibinfo {issn}
  {0301-4851}\BibitemShut {NoStop}%
\bibitem [{\citenamefont {Gagneur}\ and\ \citenamefont
  {Klamt}(2004)}]{gagneur2004}%
  \BibitemOpen
  \bibfield  {author} {\bibinfo {author} {\bibfnamefont {J.}~\bibnamefont
  {Gagneur}}\ and\ \bibinfo {author} {\bibfnamefont {S.}~\bibnamefont
  {Klamt}},\ }\href@noop {} {\bibfield  {journal} {\bibinfo  {journal} {BMC
  Bioinformatics},\ }\textbf {\bibinfo {volume} {5:175}} (\bibinfo {year}
  {2004})}\BibitemShut {NoStop}%
\bibitem [{\citenamefont {Terzer}\ and\ \citenamefont
  {Stelling}(2008)}]{terzer2008}%
  \BibitemOpen
  \bibfield  {author} {\bibinfo {author} {\bibfnamefont {M.}~\bibnamefont
  {Terzer}}\ and\ \bibinfo {author} {\bibfnamefont {J.}~\bibnamefont
  {Stelling}},\ }\Doi {10.1093/bioinformatics/btn401} {\bibfield  {journal}
  {\bibinfo  {journal} {Bioinformatics},\ }\textbf {\bibinfo {volume} {24}},\
  \bibinfo {pages} {2229} (\bibinfo {year} {2008})}\BibitemShut {NoStop}%
\bibitem [{\citenamefont {Jevremovi\'{c}a}\ \emph {et~al.}(2011)\citenamefont
  {Jevremovi\'{c}a}, \citenamefont {Trinh}, \citenamefont {Srienc},
  \citenamefont {Sosad},\ and\ \citenamefont {Daniel}}]{jevremovic2011}%
  \BibitemOpen
  \bibfield  {author} {\bibinfo {author} {\bibfnamefont {D.}~\bibnamefont
  {Jevremovi\'{c}a}}, \bibinfo {author} {\bibfnamefont {C.~T.}\ \bibnamefont
  {Trinh}}, \bibinfo {author} {\bibfnamefont {F.}~\bibnamefont {Srienc}},
  \bibinfo {author} {\bibfnamefont {C.~P.}\ \bibnamefont {Sosad}}, \ and\
  \bibinfo {author} {\bibfnamefont {B.}~\bibnamefont {Daniel}},\ }\href@noop {}
  {\bibfield  {journal} {\bibinfo  {journal} {Parallel Computing},\ }\textbf
  {\bibinfo {volume} {37}},\ \bibinfo {pages} {261 } (\bibinfo {year}
  {2011})}\BibitemShut {NoStop}%
\bibitem [{\citenamefont {von Kamp}\ and\ \citenamefont
  {Schuster}(2006)}]{kamp2006}%
  \BibitemOpen
  \bibfield  {author} {\bibinfo {author} {\bibfnamefont {A.}~\bibnamefont {von
  Kamp}}\ and\ \bibinfo {author} {\bibfnamefont {S.}~\bibnamefont {Schuster}},\
  }\href@noop {} {\bibfield  {journal} {\bibinfo  {journal} {Bioinformatics
  Application Note},\ }\textbf {\bibinfo {volume} {22}},\ \bibinfo {pages}
  {1930} (\bibinfo {year} {2006})}\BibitemShut {NoStop}%
\bibitem [{\citenamefont {Klamt}\ \emph {et~al.}(2007)\citenamefont {Klamt},
  \citenamefont {Saez-Rodriguez},\ and\ \citenamefont {Gilles}}]{klamt2007}%
  \BibitemOpen
  \bibfield  {author} {\bibinfo {author} {\bibfnamefont {S.}~\bibnamefont
  {Klamt}}, \bibinfo {author} {\bibfnamefont {J.}~\bibnamefont
  {Saez-Rodriguez}}, \ and\ \bibinfo {author} {\bibfnamefont {E.~D.}\
  \bibnamefont {Gilles}},\ }\href@noop {} {\bibfield  {journal} {\bibinfo
  {journal} {BMC Systems Biology},\ }\textbf {\bibinfo {volume} {1:2}}
  (\bibinfo {year} {2007})}\BibitemShut {NoStop}%
\bibitem [{\citenamefont {{Open Source Initiative}}(2012)}]{osi}%
  \BibitemOpen
  \bibfield  {author} {\bibinfo {author} {\bibnamefont {{Open Source
  Initiative}}},\ }\href@noop {} {\enquote {\bibinfo {title} {{The BSD
  License}},}\ }\bibinfo {address}
  {http://www.opensource.org/licenses/BSD-2-Clause} (\bibinfo {year}
  {2012})\BibitemShut {NoStop}%
\bibitem [{\citenamefont {{ETH Zurich, Computational Systems Biology
  Group}}(2012)}]{efmtoolETH}%
  \BibitemOpen
  \bibfield  {author} {\bibinfo {author} {\bibnamefont {{ETH Zurich,
  Computational Systems Biology Group}}},\ }\href@noop {} {\enquote {\bibinfo
  {title} {{efmtool - Elementary Flux Mode Tool}},}\ }\bibinfo {address}
  {http://www.csb.ethz.ch/tools/efmtool} (\bibinfo {year} {2012})\BibitemShut
  {NoStop}%
\bibitem [{\citenamefont {Orth}\ \emph {et~al.}(2009)\citenamefont {Orth},
  \citenamefont {Fleming},\ and\ \citenamefont {Palsson}}]{orth2009}%
  \BibitemOpen
  \bibfield  {author} {\bibinfo {author} {\bibfnamefont {J.~D.}\ \bibnamefont
  {Orth}}, \bibinfo {author} {\bibfnamefont {R.~M.~T.}\ \bibnamefont
  {Fleming}}, \ and\ \bibinfo {author} {\bibfnamefont {B.~O.}\ \bibnamefont
  {Palsson}},\ }\Doi {10.1128/ecosal.10.2.1} { (\bibinfo {year} {2009})},\ \doi
  {10.1128/ecosal.10.2.1}\BibitemShut {NoStop}%
\bibitem [{\citenamefont {Fukuda}\ and\ \citenamefont
  {Prodon}(1996)}]{fukuda1996}%
  \BibitemOpen
  \bibfield  {author} {\bibinfo {author} {\bibfnamefont {K.}~\bibnamefont
  {Fukuda}}\ and\ \bibinfo {author} {\bibfnamefont {A.}~\bibnamefont
  {Prodon}},\ }\href@noop {} {\bibfield  {journal} {\bibinfo  {journal}
  {Combinatorics and Computer Science},\ }\textbf {\bibinfo {volume} {1120}},\
  \bibinfo {pages} {91} (\bibinfo {year} {1996})}\BibitemShut {NoStop}%
\bibitem [{Note1()}]{Note1}%
  \BibitemOpen
  \bibinfo {note} {Note that {\protect \it efmtool} uses an inverse logic where
  0 stands for flux carrying reactions and 1 stands for not carrying a flux.
  Hence, the involved bitwise operations and comparison have to be changed
  accordingly.}\BibitemShut {Stop}%
\bibitem [{\citenamefont {{University of California, San Diego, Systems Biology
  Research Group}}(2012)}]{EcoliCoreUCSanDiego}%
  \BibitemOpen
  \bibfield  {author} {\bibinfo {author} {\bibnamefont {{University of
  California, San Diego, Systems Biology Research Group}}},\ }\href@noop {}
  {\enquote {\bibinfo {title} {{The Core E. Coli Model}},}\ }\bibinfo {address}
  {http://gcrg.ucsd.edu/Downloads/EcoliCore} (\bibinfo {year}
  {2012})\BibitemShut {NoStop}%
\end{thebibliography}%

\end{document}